\documentclass[preprint,aps,superscriptaddress]{revtex4}
\usepackage{graphicx,epsfig,amssymb,amsmath,array}
\usepackage{relsize,placeins,float}
\usepackage[dvipsnames]{xcolor}
\usepackage{blindtext}
\usepackage[english]{babel}

\newtheorem{theorem}{Theorem}

\begin{document}

\title{Orbit quantization in a retarded harmonic oscillator}

\author{Alvaro G. Lopez}
\affiliation{Nonlinear Dynamics, Chaos and Complex Systems Group, Departamento de F\'{i}sica,
Universidad Rey Juan Carlos, Tulip\'{a}n s/n, 28933 M\'{o}stoles, Madrid, Spain}

\date{\today}

\begin{abstract}
We study the dynamics of a damped harmonic oscillator in the presence of a retarded potential with state-dependent time-delayed feedback. In the limit of small time-delays, we show that the oscillator is equivalent to a Li\'enard system. This allows us to analytically predict the value of the first Hopf bifurcation, unleashing a self-oscillatory motion. We compute bifurcation diagrams for several model parameter values and analyse multistable domains in detail. Using the Lyapunov energy function, two well-resolved energy levels represented by two coexisting stable limit cycles are discerned. Further exploration of the parameter space reveals the existence of a superposition limit cycle, encompassing two degenerate coexisting limit cycles at the fundamental energy level. When the system is driven very far from equilibrium, a multiscale strange attractor displaying intrinsic and robust intermittency is uncovered. 

\end{abstract}

\maketitle

\section{Introduction}

The importance of time-delayed feedback has been extensively confirmed across many disciplines in science, ranging from mechanical physical systems \cite{ai830}, to chemical complex reactions \cite{sc86}, or complex biological systems, as for example cardiac oscillations \cite{mac77}, the propagation of impulses through the nervous system \cite{han22} or the modelling of the cell cycle \cite{fer11}. Time delays are also inescapable to understand climate phenomena, such as El Ni\~no-Southern Oscillation \cite{bou07}. In epidemiology and population dynamics retardations can be crucial as well \cite{sal98}, just as much as they are in the modelling of economic cycles \cite{kal35} or transmission lines \cite{kol86}. Indeed, whenever the forces between two physical interacting bodies are mediated by a medium or a field, or whenever large causal chains in a network of connections are reduced in a model, time-delays must be present. Consequently, the existence of retardations in differential equations describing the evolution of dynamical systems should be taken more as the rule, than as the exception.

However, this contrasts with the standard practice, where ordinary differential equations are much preferred for their simplicity, both from an analytical and a numerical point of view. Furthermore, not much attention has been dedicated to study dynamical systems where the the time retardation is state-dependent \cite{ins07,raj15,mar15,lop20}, specially in the field of fundamental physics.  Recent findings in the study of extended electrodynamic bodies using the retarded Li\'enard-Wiechert potential have shown that these particles can experience nonlinear oscillations due to self-forces, with a frequency similar to the ``zitterbewegung" frequency \cite{lop20,lop21}. The crucial importance of time-delay in atomic physics had initially been stressed by C. K. Raju \cite{raj04}. Later on, the expression ``Atiyah's hypothesis" was coined, after Sir Michael Atiyah claimed the necessity of functional differential equations to faithfully represent the dynamics of microscopic bodies, in a lecture entitled ``The nature of space", which was the first annual Einstein Public Lecture, delivered in 2005 \cite{joh06}. 

By the same year, Yves Couder and his collaborators demonstrated empirically the potential of hydrodynamic quantum analogs consisting of silicone oil droplets bouncing on a vibrating bath to describe quantum phenomena \cite{cou05,pro06}. These pilot-wave mechanical systems present striking similarities with the quantum mechanics of electromagnetically charged bodies, such as orbit quantization \cite{for10}, diffraction and interference phenomena through slits \cite{puc18}, tunneling over barriers \cite{edd09}, or the entanglement of particles \cite{pap22}. In mathematical models where the fluid is not explicitly represented \cite{bus18}, these features translate into a time-delayed feedback arising from the self-affection of the particle through the fluid medium. Just as it occurs in electrodynamics, perturbations produced in the past by the particle can affect it at the present time, introducing memory effects that can trigger its self-propulsion \cite{bus18}.

In the present work we demonstrate that the phase space orbits of a harmonic oscillator with state-dependent time-delayed feedback are quantized and organized conforming a two-level system. We also report new compelling dynamical phenomena, as for example the superposition of quantized orbits and the existence of robust intermittency. The paper is organized as follows. In Sec. 2 we introduce the mathematical model of our oscillator and explain its origin. Then, in Sec. 3 we analytically bridge state-dependent time-delayed oscillators and Li\'enard systems, proving that the periodic motion corresponds to a self-oscillation \cite{jen13}. In the following section we show that there exist quantized orbits, which are well-resolved in the energy landscape given by the harmonic external potential. Secs. 4 and 5 are dedicated to introduce two new phenomena, which might be crucial to understand other aspects of microscopic physics, such as the superposition of orbits and the passage of particles over external potential barriers. Finally, in the conclusions, we summarize the main results of the present work and the future perspectives, as usual. 

\section{Model}

We use an apparently simple model consisting in a harmonic oscillator with linear damping, according to Stokes's law of dissipation \cite{sto51}, an external quadratic potential $V(x)$ representing Hooke's law and another quadratic potential with state-dependent time-delay $Q(x_\tau)$, where $x_{\tau} \equiv x(t-\tau(x))$. Following traditional studies in classical electrodynamics, we borrow the concept of retarded potential \cite{lie98} hereafter to denote this self-excited contribution. Therefore, we can write our dynamical equation of motion in the form 
\begin{equation}
m \Ddot{x} + \mu  \dot{x} + \dfrac{d V}{d x} + \dfrac{d Q}{d x_{\tau}}=0, 
\end{equation}
where $m$ is the mass of the oscillator and $\mu$ represents the rate of dissipation.

This model is a simplified version of an oscillator recently encountered in the study of the dynamics of extended electromagnetic bodies, for which the presence of self-forces produces self-oscillations through a Hopf bifurcation \cite{lop20,lop21}. Thus, if desired, we can physically interpret to some extent this new time-delay term as the result of some complicated mass self-interactions in a mass-spring system, arising from the mass' structure. A similar, though more sophisticated, model has been previously used in the literature to study the effect of state-dependence of the delay on the phenomenon of vibrational resonance \cite{raj15}.  In summary, we can mathematically express the external potential as $V(x)=k x^2/2$ and the same holds for $Q(x_{\tau})=\alpha x_{\tau}^2/2$, yielding the differential equation
\begin{equation}
m \Ddot{x}(t) + \mu \dot{x}(t) + k x(t) + \alpha x(t-\tau(x))=0.
\label{eq:2}
\end{equation}
For convenience and without loss of generality, we shall consider $m=1$, $k=1$ and $\mu=0.1$ hereafter. We are intending to describe the dynamics of our oscillator in the phase space representation, as it is traditionally done in the study of nonlinear dynamical systems, specially regarding mechanical and electronic oscillators. However, we notice that, rigorously speaking, the true phase space of our dynamical systems is infinite-dimensional, since history functions have to be provided to integrate the Eq.~\eqref{eq:2}, instead of mere initial conditions \cite{daz17}. In the phase space $(x,y)$ we can write the differential equation as follows
\begin{align}
\dot{x} & = y \\
\dot{y} & = - 0.1 y - x - \alpha x_{\tau}.       
\end{align}

One of the crucial issues of the present work is the nature of the function $\tau(x)$. Some constraints on this function to ensure that the system is well-behaved must be provided. For example, we want the trajectories to remain bounded in the external well for $x \rightarrow \pm \infty$, so that the state-dependent delay decays to zero asymptotically. It is also reasonable to demand that the delay function $\tau(x)$ remains bounded all over its domain, guaranteeing that the feedback coming from the past history of the dynamics does not extend to minus infinity. In this manner, we bound the memory of this non-Markovian system to a finite domain of its temporal past. Finally, symmetry with respect to spatial reflections ($x \rightarrow - x$) is also present in the original model \cite{lop20}. Moreover, as we show ahead, we can exploit the degeneracy introduced by this symmetry to obtain intriguing new dynamical phenomena. A simple function that has been used in previous works is the Gaussian distribution \cite{raj15}, which accounts for these three requirements. In conclusion, we assume that $\tau(x)=\tau_0 e^{-x^2/2 \sigma^2}$, and fix $\sigma=1/\sqrt{2}$ unless otherwise stated. The parameter $\tau_0$ represents the maximum value of the time-delay feedback, attained at the centre of the potential well. It consitutes one of the two key parameters investigated in the present study. 

All things considered, we have a time-delayed nonlinear oscillator with two independent parameters $\alpha$ and $\tau_0$. Interestingly, we note that the system's nonlinearity comes entirely from the retardation, since both potentials have been assumed harmonic. Even though this system has been designed following previous findings in electrodynamics, we would like to stress all the simplifications performed. Firstly, the delay in the functional differential equation appearing in Ref.~\cite{lop20} depends both on the speed and the acceleration of the particle. Secondly, such differential equation is of the advanced type \cite{lop20}, since the acceleration and the speed appearing in the Li\'enard-Wiechert potentials are retarded themselves. Unfortunately, numerical schemes to integrate advanced differential equations with state-dependent delays are lacking. This has motivated the authors to develop the present approximated model. Finally, some speed-dependent nonlinearities appearing in the dissipation term and also in the restoring force term have been neglected. They are related to the Lorentz's gamma factor, which is required to comply with the principle of relativity in classical electrodynamics. Consequently, the present model remains somewhat abstract. It is not our purpose to rigorously fit it to any specific physical system. We just use it to illustrate some physical phenomena that are frequently believed to belong exclusively to the atomic realm of physics.

\section{Related Li\'{e}nard system}

Given the fact that there is dissipation in the system, in the absence of retardation ($\alpha=0$ or $\tau_0=0$), it can be immediately proved using the Lyapunov energy function $E(x,y)=(x^2+y^2)/2$ that the rest state at the equilibrium $x=0$ is the only global stable fixed point of the system, which asymptotically attracts all the initial conditions in the phase space \cite{ali96}. We recall that this function only comprises the conservative part of the energetic content of our dynamical system. However, when the retarded potential is activated for $\alpha>0$, as we increase $\tau_0$ bellow a critical value, a Hopf bifurcation appears destabilizing such an equilibrium point. A fundamental energy level appears with non-zero energy fluctuations, in which the system performs limit cycle oscillations. Therefore, the orbit becomes quantized as a consequence of the time-delayed feedback. The system becomes unstable and locally active \cite{mai13}, performing a periodic self-oscillatory motion around the minimum of the square well potential. Of course, this is only possible at the expense of an energy input in the system, which must come from external field sources \cite{jen13,lop20}. Therefore, the present dynamical system must be regarded as as a non-equilibrium open physical system, whose nonlinear periodic motion can be interpreted as a cyclic thermodynamic engine \cite{lop22}. Due to the existence of energy losses, these dynamical systems are frequently named dissipative structures. This contrasts to conservative dynamical systems, which are generally equipped with a symplectic structure \cite{abr08}.

We now prove that the dynamics of the system is a self-oscillation, triggered by the well-known Hopf bifurcation. To compute the value of $\tau_0$ at the bifurcation point, we approximate this system to a Li\'enard system. Expanding in Taylor series the retarded potential to second order yields
\begin{equation}
x(t-\tau(x))=x(t)-\tau(x) \dot{x}(t)+\dfrac{1}{2}\tau^2(x)\Ddot{x}(t)+O(\tau^3).  
\end{equation}
When the time-delay is small, we can neglect the third and higher order terms, substitute the two leading order contributions in the Eq.~\ref{eq:2}, and obtain
\begin{equation}
\Ddot{x} + f(x)\dot{x} + g(x) = 0,
\label{eq:6}
\end{equation}
where the functions $f(x)=(\mu-\alpha \tau(x))/(1+\alpha \tau^2(x)/2)$ and $g(x)=(k+\alpha)/(1+\alpha \tau^2(x)/2)$ have been defined. Thus, as we can see, small retardations have two fundamental physical consequences. Firstly, an antidamping correction to the drag force appears to first order. Secondly, the inertia of the mass becomes dependent on the dynamical state through its evolution along the trajectory. To demonstrate that this Li\'enard system with $f(x)$ and $g(x)$ as defined, fulfils the conditions required to produce the Hopf bifurcation, we appeal to Li\'{e}nard's theorem. Given the importance of this theorem, we first enunciate it, so that the reader is aware of all the technical details \cite{lie28}. For this purpose it is convenient to introduce the primitive function $F(x)=\int^{x}_{0} f(s)d s$, since it is alluded in the theorem, which reads

\begin{theorem}[Li\'enard, 1928]
Under the assumptions that the functions $F(x),g(x) \in \mathcal{C}^1(\mathbb{R})$ are odd, $x g(x)>0 $ for $x \neq 0$, $F(0)=0$, $F'(0)<0$ and $F$ has one single root for $x=a$, beyond which it increases monotonically to infinity, it follows that there exists only one limit cycle and that it is stable.  
\end{theorem}

\begin{figure}
\centering
\includegraphics[width=1.0\linewidth,height=0.5\linewidth]{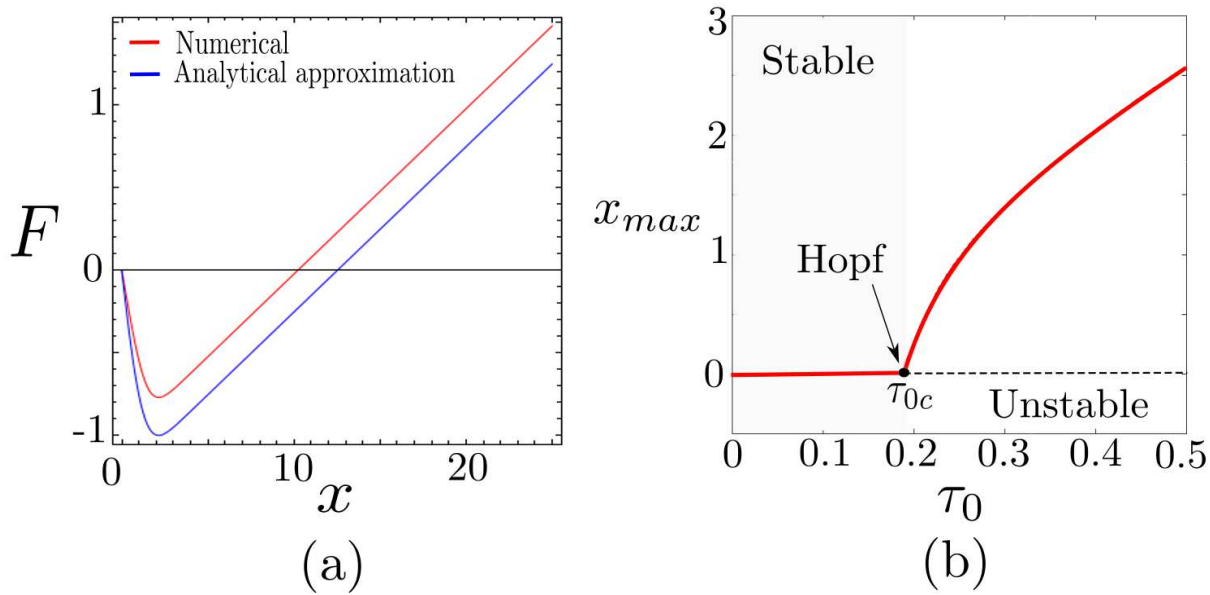}
\caption{\textbf{Hopf bifurcation}. The conditions for Li\'enard's theorem are shown. (a) The function $F$ for $\tau_0=1$ and $\alpha=1/2$. It clearly verifies $F(0)=0$, $F'(0)<0$, has a single root for $a \approx 10$ and increases monotonically without bounds thereafter. An analytical approximation using the error function. (b) The bifurcation diagram, with the maximum value of $x$ along the orbit resulting from the numerical solution of the retarded differential (red curve), showing a Hopf bifurcation for the value $\tau_{0c}=0.195$, very close to the analytical prediction of the related Li\'{e}nard system, which corresponds to $\tau_{0c}=0.2$. The rest state becomes unstable beyond the critical bifurcation point, entailing the zero-point fluctuations of the fundamental energy level.}
\label{fig:1}
\end{figure}

For a proof of the theorem we refer the reader to Ref.~\cite{per91}. It is immediate to confirm that all the requirements for the existence of the Hopf bifurcation are accomplished. Indeed, the function $F$ fulfils $F(0)=0$, has a root at $a \in \mathbb{R}^{+}$, and for $x \geq a$ we find that it increases monotonically towards infinity (see Fig.~\ref{fig:1}(a)). This follows from an analytical estimation of $F(x)$, which can be provided by neglecting the postive term $\tau^2(x) \ll 2/\alpha$ in the denominator of $f(x)$, yielding the function $F(x) = \mu x - \sqrt{2 \pi} \alpha \tau_0 ~\text{erf}(x/\sqrt{2})$, with $\text{erf}(x)$ the error function. It is also confirmed by the numerical solution of the integral, which has been computed using the trapezoidal rule. Recall, this rule works by approximating the region under the graph of a function as a trapezoid and calculating its area. The condition $x g(x)>0$ is trivially verified, while the condition $F'(0)>0$ can be used to find out the value of the Hopf bifurcation, since $F'(0)=(\mu-\alpha \tau_0)/(1+\alpha \tau^2_0/2)$, which entails that $\tau_0>\mu/\alpha$. For the parameter values here considered and $\alpha=1/2$, we can approximate the value of the point where the Hopf bifurcation takes place to $\tau_0=1/5$. This analytical results holds nicely, as depicted in Fig.~\ref{fig:1}(b). In the following section we show that, as we push the system even further from thermodynamic equilibrium \cite{lop22} by increasing the effects of the time-delay feedback, the oscillator undergoes further bifurcation phenomena producing a second quantized excited orbit, at a higher energy level. 

\section{Energy levels: multistability}

Once we have demonstrated that the time retardation can destabilize the rest state, generating a fundamental energy level with zero-point fluctuations, it is worth asking if by posing the system even further from equilibrium, \emph{i.e.} by increasing the delay feedback, orbits with larger amplitude representing excited energy levels can appear. For this purpose we have computed the bifurcation diagrams of the related maxima map of the system. As it is well-known, this map can be constructed by computing the local maxima of the temporal series. Together with its related minima map, this is the simplest general way to discretize the dynamics of delayed differential equations. We insist again that, strictly speaking, the phase space of retarded differential equations is infinite-dimensional. An alternative possibility is to build an embedding from the temporal series and to construct a Poincar\'e section out of it. However, this technique it computationally more intensive and does not produce better insights into the dynamics of the system. 

To compute the bifurcation diagrams we have to integrate the Eq.~\eqref{eq:2}. This requires to consider history functions \cite{daz17}. Since in the absence of time retardation the system is harmonic, we consider that the most natural choice of history functions are periodic solutions, Therefore, we take the functions $x(t)=A \sin(\omega t + \varphi)$ for $t<0$. Moreover, this choice can be used later to ascertain relevant dynamical aspects of the system under finite-time external periodic drivings, which can be physically interpreted as brief pulses exerted on the oscillator. As it is expected, external perturbations acting on the system can produce transitions between the energy levels. 

\begin{figure}
\centering
\includegraphics[width=1.04\linewidth,height=0.35\linewidth]{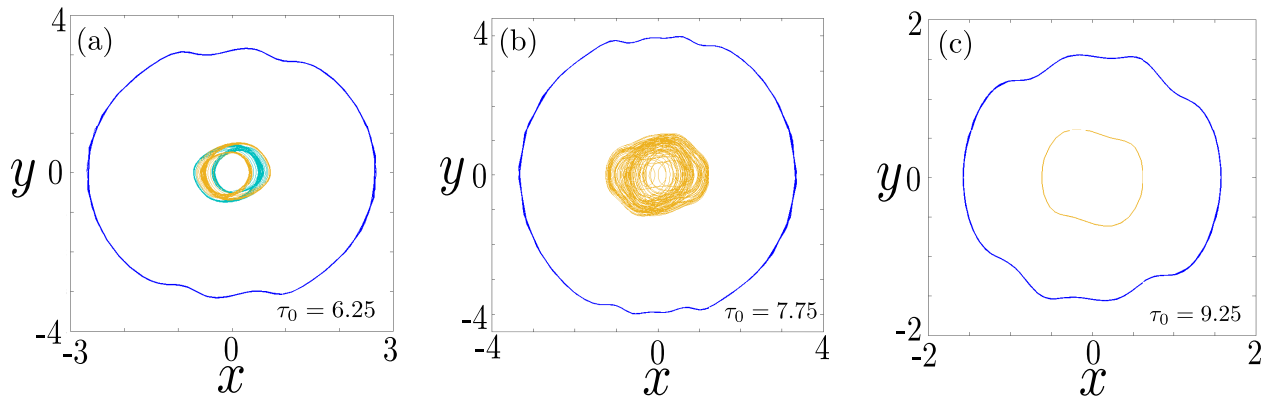}
\caption{\textbf{Multistability}. Three phase space portraits along a quasiperiodic route to chaos in the multistable region for $\alpha=0.5$ (see first bifurcation diagram in Fig. 2). (a) Two doubly-degenerate quasiperiodic attractors coexist with a higher amplitude periodic limit cycle surrounding them. (b) The quasiperiodic attractors have merged into a single chaotic attractor as the delay increases, while the most exterior limit cycle has enlarged. (c) The chaotic orbit disappears through a crisis for even higher time-delays, yielding two coexisting periodic limit cycles of different amplitude.}
\label{fig:3}
\end{figure}

Because our aim is to figure out if there exists multistable parameter regimes, represented by two or more coexisting stable limit cycles, we throw ten different initial conditions randomly chosen in the range $A \in [0,3]$, $\omega \in [-\pi,\pi]$ and $\varphi \in [0, \pi]$. We compute the trajectories in the temporal interval $t \in [0, 2000]$ using a residual order integrator implemented in MATLAB. Transients as long as seven-tenths or even larger of the whole temporal series are discarded, since time-delayed systems usually display long transient phenomena \cite{lak11}. Finally, we obtain the maxima map and represent these points for $1200$ varying parameter values of the maximum time-delay in the range $\tau_0 \in [0,~11]$. Recalling that several conditions can lead to the same asymptotic limit cycle, we have coloured the bifurcations diagrams in two colors, to clearly distinguish the two energy levels, whenever they exist.
\begin{figure}
\centering
\includegraphics[width=1.0\linewidth,height=0.5\linewidth]{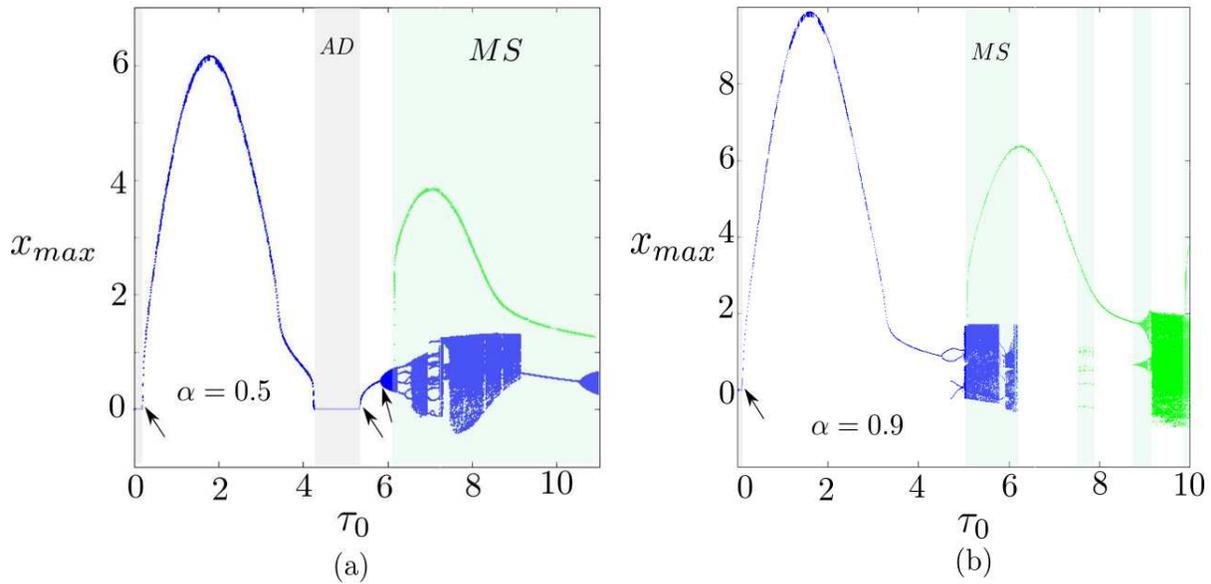}
\caption{\textbf{Bifurcation diagrams ($\alpha>0$)}. The bifurcation diagrams of the maxima map of $x$ are represented for increasing values of the maximum delay $\tau_0$. A total number of ten initial randomly chosen histories have been used, and depicted using two different colors to represent the asymptotic sets, clearly distinguishing multistable regions (green background). (a) For $\alpha=0.5$ several Hopf bifurcations (arrows), interrupted by an amplitude death region (AD), end in a quasiperiodic route to chaos. Multistability (MS) starts at the critical value $\tau_{0c}=6.1$, when a high energy limit cycle (green) is born, coexisting with the fundamental energy level, which involves periodic, quasiperiodic or chaotic attractors (blue). (b) For $\alpha=0.9$ similar results are observed, except for the disappearance of the amplitude death region, and the fact that the mustilstable regions appear and disappear through several crises.}
\label{fig:2}
\end{figure}
As we can see in Fig.~\ref{fig:2}(a), and as detailed in the previous section, for $\alpha=1/2$, as the time-delay is increased from zero, a first Hopf bifurcation reveals at $\tau_0=1/5$. Then, if we increase further the maximum delay $\tau_0$, the fundamental orbit first enlarges reaching a maximum amplitude of $x_{max}=6.0$ for $\tau_0$ close to $2.0$, then shrinks again and, finally, it disappears. This is the well-studied phenomena of amplitude death (AD), frequently displayed by time-delayed differential equations \cite{ram98}. However, for values beyond $\tau_0=5.25$ a Hopf bifurcation shows anew, which is now followed by a secondary Hopf bifurcation, giving rise to quasiperiodic motion. For an approximate critical value of the maximum time-delay $\tau_{0c}=6.1$, a new periodic limit cycle of higher amplitude is born, rendering a multistable (MS) two-level system. The second quantized excited state shall persist all along the bifurcation diagram and remains periodic, although its amplitude shrinks as the retardation increases. Then, the fundamental energy level experiences further bifurcations through a quasiperiodic route to chaos \cite{spr08}, ending in a chaotic strange attractor (see Figs.~\ref{fig:3}(a) and (b)). The chaotic attractor experiences a crisis at approximately $\tau_{0}=9.0$, yielding two coexisting periodic limit cycles, which are depicted in Fig.~\ref{fig:3}(c). For $\alpha=0.9$ similar results are observed in Fig.~\ref{fig:2}(b), except for the fact that the amplitude death region is missing, and also the multistable regions appear and disappear intermittently along the bifurcation diagram through several crises. Some new interesting dynamical features are also discerned, as for example the coexistence of two quasiperiodic attractors for $\tau_0=9.5$. Naturally, whenever a strange attractor disappears through a crisis, transient chaos phenomena \cite{tel06} can be observed, where a trajectory can spend large transients in the fundamental level, and then spiral away towards the first excited level.

\begin{figure}
\centering
\includegraphics[width=0.95\linewidth,height=0.5\linewidth]{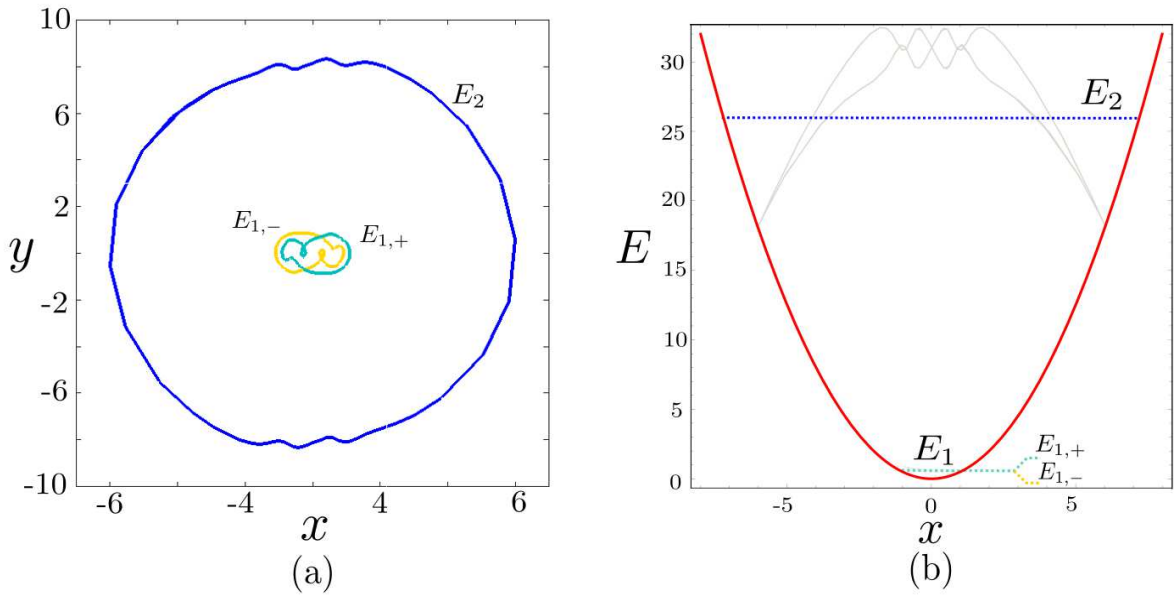}
\caption{\textbf{Energy levels}. A two-level system for $\alpha=0.9$ and $\tau_0=5.87$. The fundamental level $E_1$ is doubly degenerate, with two coexisting symmetric (under reflection) limit cycles, $E_{1,+}$ and $E_{1,-}$. (a) Limit cycles representing the quantization of orbits, with two different average energies, one corresponding to the fundamental level, and the other to the first and last excited level. (b) The harmonic potential is represented in red, while the average energy of the limit cycles is represented with dashed lines. In gray we can see the detour of the orbit through different energy shells. The fluctuations are considerable, although the two levels are well resolved.}
\label{fig:4}
\end{figure}

We now investigate if the two energy levels are well-resolved across the different energy shells. For this purpose, and also for aesthetic purposes, we have used a value of $\alpha=0.9$ and $\tau_0=5.87$ to illustrate this two-level system. For these parameter values, we can find two stable symmetric degenerate coexisting orbits at the fundamental level, as shown in Fig.~\ref{fig:4}(a). This degeneracy is a consequence of the fact that the Eq.~\eqref{eq:2} is invariant under spatial reflections, and the splitting of these two orbits constitutes a typical phenomenon of symmetry breaking at the fundamental energy level. We recall that symmetry breaking is an ordinary phenomenon frequently observed in nonlinear self-excited systems \cite{mai13}. In Fig.~\ref{fig:4}(b) we have plotted the harmonic external potential in red. We have used the Lyapunov energy function $E(x,y)=(x^2+y^2)/2$ to compute the energy of the particle along the limit cycles \cite{per91}, and numerically integrated its average value along these periodic orbits, using the trapezoidal rule once more. The average energy has been plotted in the energy diagram in dashed lines, together with the energy fluctuations that the stable quantized orbits experience along their periodic motions. As we can clearly appreciate, despite the fact that the fluctuations are substantial and the oscillator performs excursions out of shell with respect the average energy, the two levels are well differentiated and they do not overlap in the energy diagram. Consequently, it can be safely stated that the present system displays quantized stable orbits at two independent energies, which can be denoted as $E_1=0.4$ and $E_2=26$.

\begin{figure}
\centering
\includegraphics[width=0.95\linewidth,height=0.43\linewidth]{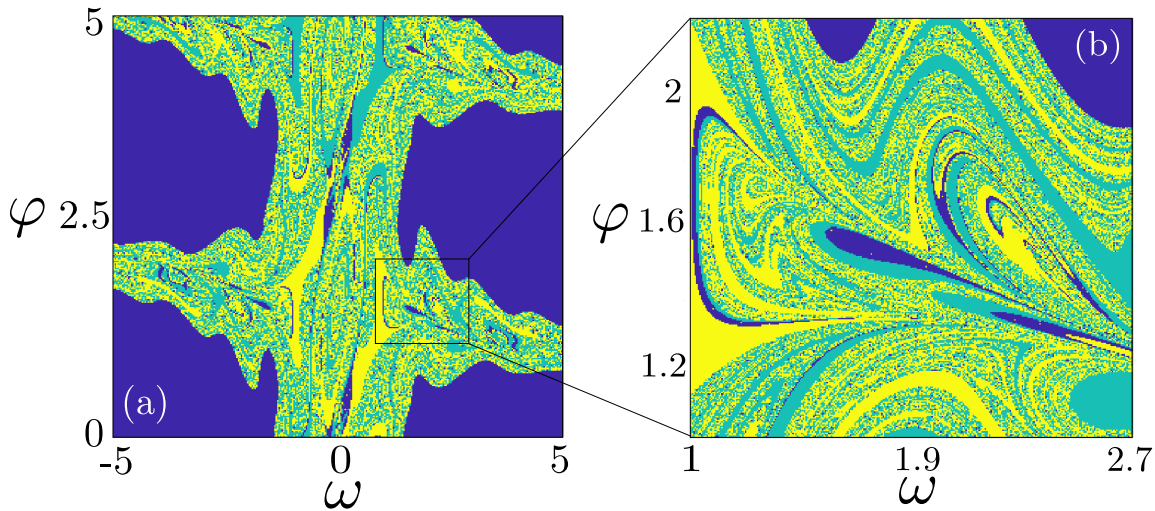}
\caption{\textbf{Unpredictability}. The basins of attraction in the history space of the three stable attractive orbits for $\alpha=0.9$ and $\tau_0=5.87$. The two energy levels are clearly mixed in the phase space of initial histories, rendering the basins their fractal nature. Thus, arbitrarily small perturbations in the initial histories can lead to different asymptotic energy levels. (a) The basin of attraction for $A=0.43$ and varying frequency in the phase space of the periodic histories. (b) A blow-up of the basins, evincing the sensitivity of the system to initial conditions, which entails unpredictability at all scales of precision.}
\label{fig:5}
\end{figure}

To conclude this section, we have also studied the basins of attraction of the system for this particular situation, to ascertain if there exists sensitivity to external perturbations. This is of crucial importance, for if an external perturbation is effected on this system, we may wonder which of the possible asymptotic limit cycles is attained in the end. Or, equivalently, we may ask about the ultimate energy of the oscillator when it is perturbed from the outside. In Fig.~\ref{fig:5} we show the basins of attraction in the history subspace of periodic functions. We have used a resolution of $300 \times 300$, fixed an amplitude of $A=0.43$, and computed trajectories until they get close enough to one of the three attractors. Depending on which attractor is approached, each initial history is plotted in the parameter space with a different color. As we can see, the basins are fractalized, what introduces unpredictability at all the scales of precision  \cite{agu09}. However, this basin does not posses the Wada property \cite{daz17}. In general, unless infinite experimental accuracy is accessible, the best that we can say is that there exists some probability that the system might end in one of the two energy levels. This probability can be roughly approximated by merging the two basins of the respective orbits at the fundamental level, and by computing the size of the resulting basins of attraction in the parameter space.  The fraction of volume of each basin in relation to the total volume in the parameter space in the region at investigation allows to introduce the concept of basin stability \cite{men13}. In addition, the asymptotic uncertainty can be further studied through the concept of basin entropy, which offers a more concise probabilistic account of the hidden structure of the basins \cite{daz16}. 

\section{Limit cycle superposition}

The present section is dedicated to describe a new dynamical phenomenon that we have encountered for $\alpha<0$, which reminds of phenomena typically appearing in microscopic physics. In fact, the Eq.~\eqref{eq:2} with $\alpha<0$ resembles more exactly the electrodynamic self-oscillator encountered in previous works \cite{lop20}. Specifically, we refer to the existence of states of superposition of orbits. In the present case, this corresponds to a quasiperiodic limit cycle encompassing two smaller symmetric degenerate limit cycles. This phenomenon can only be detected when the effects of the retarded potential are comparable to the magnitude of the external potential. Here we have selected a value of $\alpha=-0.9$ to illustrate the phenomenon, which in absolute value is rather close to the value $k=1$. 
\begin{figure}
\centering
\includegraphics[width=0.6\linewidth,height=0.5\linewidth]{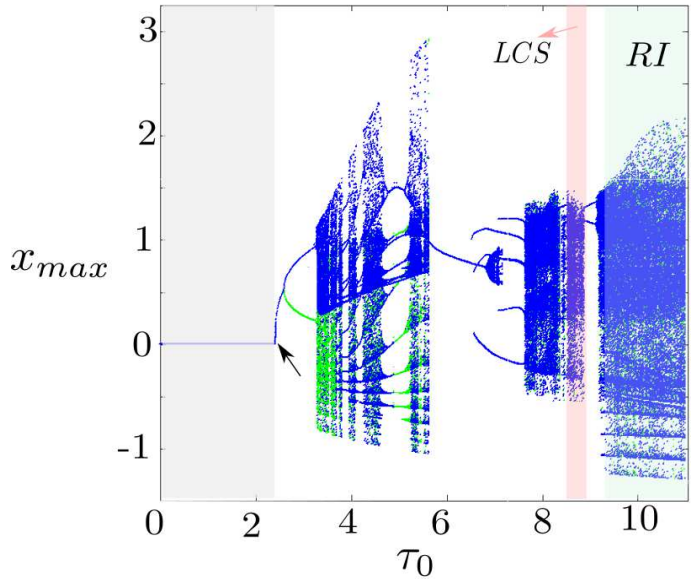}
\caption{\textbf{Bifurcation diagram ($\alpha<0$)}. The bifurcation diagrams of the maxima map of $x$ are represented for increasing values of the maximum delay $\tau_0$ and $\alpha=-0.9$. A total number of ten initial randomly chosen histories have been used and depicted using two different colors to represent the asymptotic sets. A first Hopf bifurcation (arrow) appears now for $\tau_0=2.4$, followed by a Pitchfork bifurcation (green and blue branches). We can distinguish a region where limit cycle superposition (LCS) is detected (red background). For $\tau_0>9.2$ a strange attractor with robust intermittency (RI) appears.} 
\label{fig:6}
\end{figure}

In the first place we plot the bifurcation diagram. It has been computed following exactly the same recipe described in the previous section. As the reader can see in Fig.~\ref{fig:6}, for $\alpha<0$ we cannot find a corresponding Li\'enard system that experiences a Hopf bifurcation for small values of the maximum time-delay $\tau_0$. This occurs because the change in the sign of $\alpha$ precludes the antidamping effect produced in the first derivative of $x$ appearing in Eq.~\eqref{eq:6}. However, as $\tau_0$ is further increased, again a Hopf bifurcation reveals at the approximate maximum delay critical value $\tau_{0c}=2.4$. Thus, now, the instability occurs when the system is posed quite far from the original equilibrium. It must be the result of high-order terms in the Taylor expansion of the delayed potential, involving the jerk, the jounce and other derivatives of higher order. Later on, at the critical value $\tau_{0c}=2.7$, a Pitchfork bifurcation ensues, which then transits to the chaotic regime, as we keep increasing the retardation. As far as we have computed, a period three orbit coexisting with the two period one orbits suddenly appears. As we zoom in the bifurcation diagram, we can see that these period-3 orbits then experience a period doubling bifurcation. Nevertheless, the cascade cannot be clearly distinguished, since it includes very complicated dynamics with truly large chaotic transients, involving heterogeneous alternating motions.

\begin{figure}
\centering
\includegraphics[width=0.82\linewidth,height=0.5\linewidth]{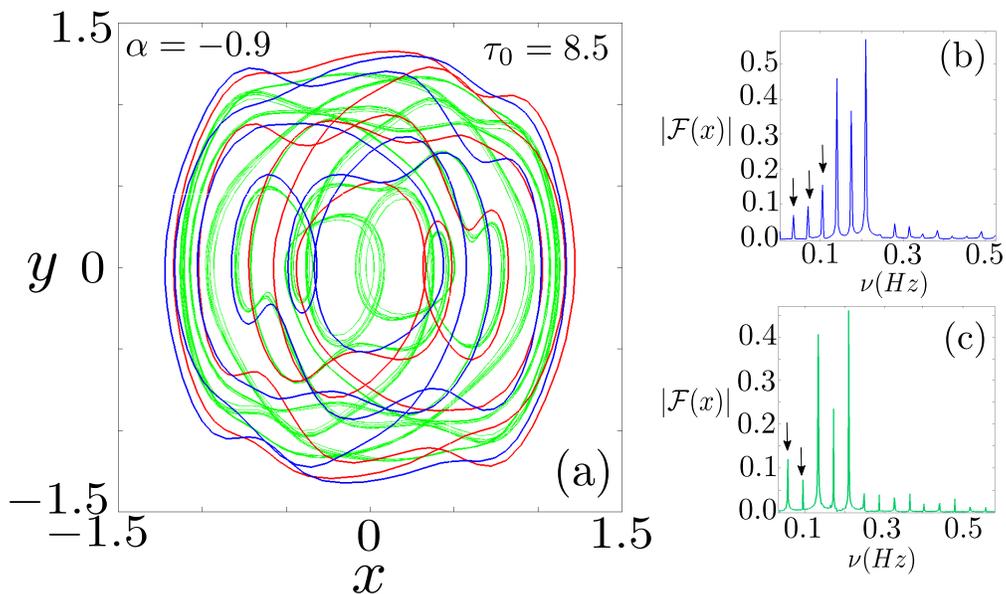}
\caption{\textbf{Limit cycle superposition}. We can see two symmetric degenerate periodic limit cycles at the fundamental (red and blue orbits) level for $\tau_0=8.5$ and $\alpha=-0.9$. Another limit cycle (green orbit) encompassing the previous two orbits can be appreciated. (b) Power spectra of the periodic orbits. (c) Power spectra of the superposition limit cycle encompassing the periodic orbits, where the lower frequencies (arrows) are different, rendering this attractor its quasiperiodic nature.}
\label{fig:7}
\end{figure}

For higher values of the maximum time-delay, around $\tau_0=8.5$, we can find a window of parameter values in the bifurcation diagram where a limit cycle superposition can be found. We describe this new phenomenon in detail. In this region, we have numerically detected at least five different coexisting limit cycles, by scrutinizing the history parameter space. Two of them are symmetric and have lower amplitude. They could describe a first fundamental level, but this time unresolved from the second, which is the one concerned now. For simplicity, we omit them from our analysis. Then, another two limit cycles of larger amplitude have also been found, which consist in two complicated period-6 stable symmetric degenerate orbits. By varying initial histories in the parameter space $(A,\omega,\varphi)$, one can find many past histories leading to any these two limit cycles, just as shown in Fig.~\ref{fig:5} for $E_{1,\pm}$. But, to our surprise, we have also found a superposition limit cycle travelling along both limit cycles (see Figs.~\ref{fig:7}(a)). This orbit spends some time going close to one of the degenerate stable periodic orbits, and then switches to the other one, alternating between them in a regular fashion. 

The new limit set corresponds to an apparently quasiperiodic stable attractor, and it can also be accessed from many parameter values $(A,\omega,\varphi)$ in the parameter space chosen as initial histories. Since this superposition limit cycle resembles to its encompassed orbits, it can be numerically shown that its average energy is, although slightly below, close to the average energy of the other two period-6 orbits. The small difference arises because the superposition limit set visits regions of the phase space with lower energy (closer to the origin of the square well), which are not covered by the periodic trajectories. Thus, as far as we are concerned, we describe here for the first time a stable limit cycle that can be partly constructed from two smaller stable orbits, by nearly taking their union in the phase space. This would be impossible in a finite-dimensional dynamical system represented by some set of ordinary differential equations, as they are frequently used to describe conventional mechanical conservative systems: two orbits cannot cross in the phase space of a finite-dimensional continuous system. Of course, if we interpret the true phase space of our retarded oscillator as infinite-dimensional, neither they do cross here.

To conclude our analysis, because the superposition state takes after the two encompassed smaller cycles, we have computed the power spectra (see Figs.~\ref{fig:7}(b) and (c)) of the temporal series of the quantized periodic orbits and their superposition orbit, to ascertain the periodicity of the later. As expected, the power spectra of both orbits take after one another, since their average energy is similar. However, we can see that differences appear in the lower frequency domain of the spectrum, which render the superposition limit cycle quasiperiodic or, in the worst case, of a very high period, as compared to the other orbits. Nevertheless, without taking advantage of spectral analysis, it is really striking to see how this quasiperiodic orbit resembles to the underlying periodic limit cycles. 
\begin{figure}
\centering
\includegraphics[width=1.00\linewidth,height=0.43\linewidth]{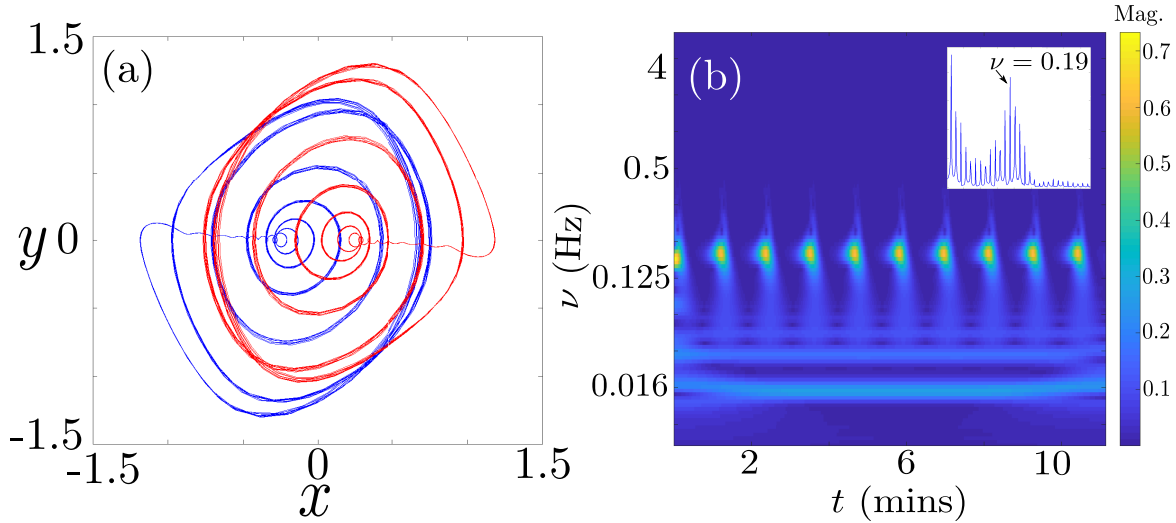}
\caption{\textbf{Multiscale limit cycles}. (a) Two degenerate symmetric limit quiasiperiodic attractors for $\alpha=-0.9$ and $\tau_0=3.35$. The trajectories are reminiscent of a saddle-focus projected on the 2D phase space. (b) The magnitude of the continuous wavelet transform is represented (colorbar), using the analytic Morse wavelet with the symmetry parameter  equal to 3 and a time-bandwidth product equal to 60. We can already appreciate in the temporal evolution of the spectrum a complex on-off periodic oscillatory behaviour. The inset shows the total power spectra, with a bimodal distribution displaying a rich frequency content.}
\label{fig:8}
\end{figure}

\section{Robust intermittency}

We now investigate an interesting dynamical phenomenon that is encountered in our retarded oscillator for $\alpha=-0.9$ when the maximum time-delay is substantially increased (see Fig.~\ref{fig:8}(a)). This phenomenon consists in a multiscale strange attractor that appears to be robust \cite{ban98} and which also exhibits intrinsic intermittency in a double sense. To understand it properly, we first show some complicated symmetric degenerate limit cycles with two intrinsic scales. By intrinsic we mean a property that results from the structure of the limit cycles, and not as a consequence of some crises at a bifurcation point. As shown in Fig.~\ref{fig:8}(a), for $\tau_0=3.35$, this attracting orbits spiral out of the rest state and then are reinjected back to the limit cycle, drifting slowly towards the equilibrium point without oscillating at all. They clearly evoke a saddle-focus structure, as appearing in Shilnikov's bifurcation \cite{shi67}, specially when embedded in a higher dimensional subspace of the full infinite-dimensional true phase space (see below). Their frequency spectrum is very rich, having two maxima and many frequencies at different scales. Interestingly, by implementing a continuous wavelet transform method, we can capture dynamical phenomena that is not displayed by conventional stationary spectral analysis. As it can be appreciated in Fig.~\ref{fig:8}(b), this time-multiscale method uses several time-windows, showing how the frequency spectrum evolves in time, and evincing the alternation in the system between oscillatory dynamics and low-speed silent drifts. This dynamics is somewhat reminiscent of relaxation oscillators, although these limit cycles are way more sophisticated in the present case \cite{van20}.
\begin{figure}
\centering
\includegraphics[width=1.04\linewidth,height=0.75\linewidth]{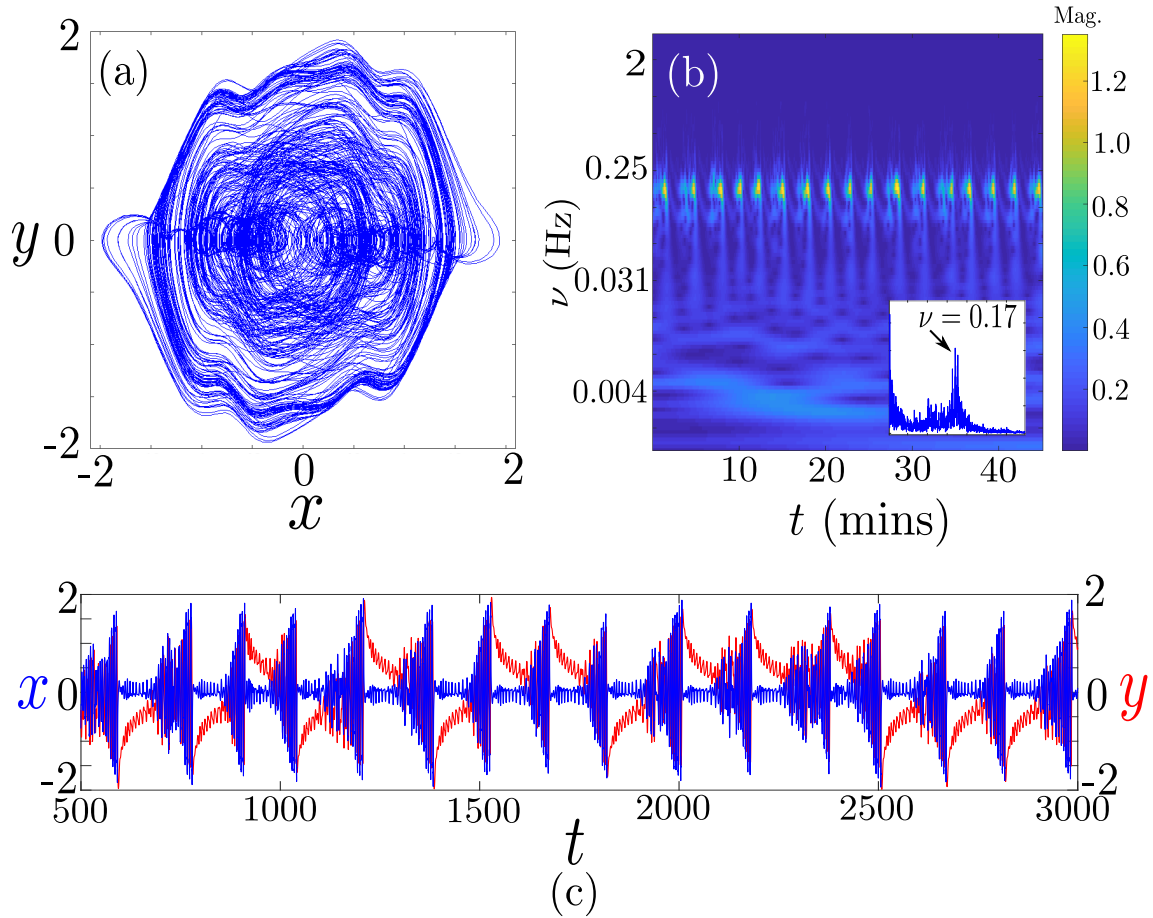}
\caption{\textbf{Intrinsic intermittency}. (a) The two degenerate symmetric limit quiasiperiodic attractors have merged into a chaotic strange attractor in the 2D phase space. This attractor possess two dynamical and well differentiated scales. (b) The continuous wavelet transform is represented (colorbar), using again the analytic Morse wavelet with the symmetry parameter equal to 3 and a time-bandwidth product equal to 60. We can newly appreciate in the temporal evolution of the spectrum a complex behaviour that switches between two oscillatory motions with different amplitude. (c) The time series of $x$ and its derivative $y$ in the phase space. A sequence of bursts is clearly appreciated. Note how the trajectories can be reinserted into the attractor through two different arms, making the phenomenon doubly intermittent.}
\label{fig:9}
\end{figure}

For higher parameter values, as for example for $\tau_0=9.5$, these two complex limit cycles have merged into a strange chaotic attractor, as shown in Fig.~\ref{fig:9}(a). Now we find that the system alternates between two different states of chaotic oscillation, one with low amplitude and another with a higher amplitude (Fig.~\ref{fig:9}(c)). In this sense, we can affirm that the system displays intermittent behaviour, switching between these two nonperiodic modes of oscillation. Comparing this dynamics with the dynamics along the underlying multiscale limit cycles previously described, we can say that the low-speed drift towards the original equilibrium of the system without retardation, have now become an oscillation of small amplitude around it. Note also how the system is reinjected into the domain through two possible routes: the lower branch and the higher branch of the residual multiscale attractors, rendering a second form of intermittency. Importantly, this doubly intermittent behavior is intrinsic to the complex heterogeneous nature of the attractor. Simply put, it does not require a fine-tuning of the parameter $\tau_0$, as opposed to conventional intermittency phenomena, which occurs close to bifurcation critical points \cite{pom80}. Moreover, it can be shown that this chaotic attractor does not disappear as we move across the parameter space $\tau_0$. Thus it is robust under parameter perturbations. 
\begin{figure}
\centering
\includegraphics[width=1.0\linewidth,height=0.4\linewidth]{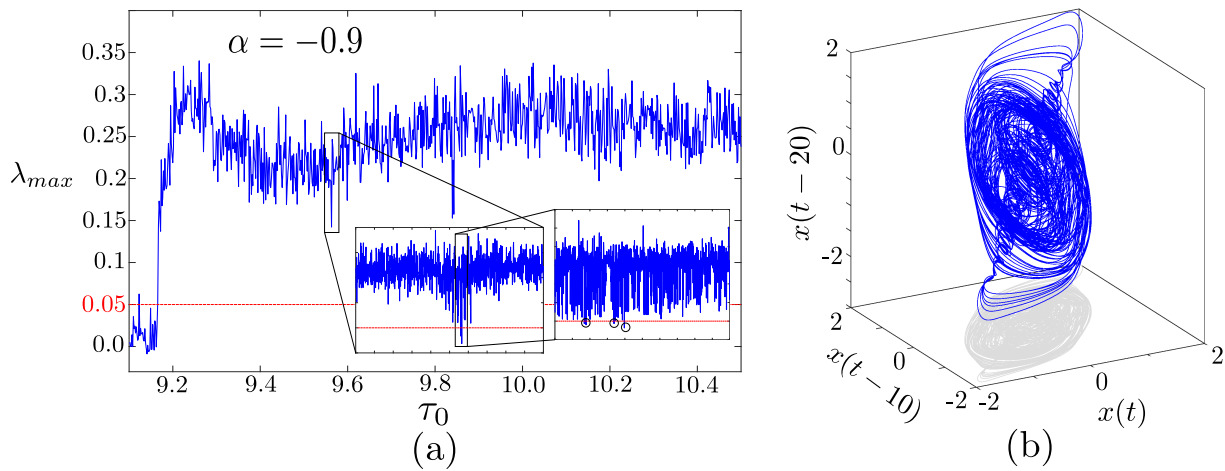}
\caption{\textbf{Robustness}. (a) The largest Lyapunov exponent has been computed by using embedding techniques across different values of the time-delay for $\alpha=-0.9$. A value of 0.05 has been chosen as a threshold to determine if the motion is chaotic. We see that its positive value rarely goes below the threshold, what entails great robustness of the attractor to parameter perturbations. (b) The attractor embedded in $D=3$ dimensions, reconstructed with an embedding delay $\tau=10$. We see how it unfolds in this higher-dimensional space, so that the saddle-focus hidden structure is more clearly appreciated. Its projected shadow manifestly resembles the attractor in 2D phase space.}
\label{fig:10}
\end{figure}

Fascinated by this dynamical behavior and by the fact that the attractor seems to be robust, in the sense that no periodic windows appear as we zoom in the bifurcation diagram around some value of $\tau_0$, we have computed the largest Lyapunov exponent (LLE) across a continuous interval of parameter values of the maximum time-delay $\tau_0$. Since MATLAB's integrator does not allow to compute the LLE dynamically, we have taken advantage of embedology and used the entire time series. We follow a method exposed by Rosenstein et al. to efficiently compute the LLE from experimental time series \cite{ros93}. These computations have been carried out using an embedding dimension of $D=3$, and embedding time-delay for the series of $\tau=10$. The mean period $T$ to compute the LLE considered can be obtained from spectral analysis (see Ref.~\cite{ros93}). We have used a value of $T=35$, which is an upper bound obtained for many parameter values of the attractor. The time of integration has been considered $t \in [0, 3000]$ and the maximum number of iteration for the algorithm was set to $1500$, keeping our conservative attitude (see again Ref.~\cite{ros93}). The 3D embedding is depicted in Fig.~\ref{fig:10}(b).

In Fig.~\ref{fig:10}(a) we can see the value of the maximum Lyapunov exponent for $\alpha=-0.9$, starting with a periodic orbit at $\tau_0=9.1$, where the value of the Lyapunov exponents is very small or negative, as it should be for a periodic stable motion. When the chaotic attractor is born, a sudden jump to positive high values of the exponent is computed. We have set a threshold of $\lambda_{max}=0.05$ as the limiting value below which we cannot safely affirm that a sensitivity to initial histories occurs. This value is a conservative choice consistent with the temporal series of the periodic window, before the chaotic dynamics is triggered. As shown in Fig.~\ref{fig:10}(a), we have performed magnifications at several scales whenever downward peak fluctuations in the LLE exponent are present. The threshold limit is rarely exceeded. Furthermore, whenever the exponent drops bellow the value of 0.05, we have systematically computed bifurcation diagrams to see if the chaotic behavior vanishes. However, we have not found any periodic windows, and if periodic orbits exist, they coexist with the chaotic attractor. Thus we can conclude that the chaotic attractor is very robust in the present dynamical system, even though an analytical proof of robustness can not be easily provided in this case, as in previous works \cite{ban98}. Since the intermittency arises as a consequence of the complicated nature of the attractor, which is robust, it is reasonable to say that, in addition to being intrinsic, it is robust, as well.

\section{Conclusions} 

In the present work we have developed a very simple retarded oscillator with state-dependent delays, uncovering crucial dynamical behaviour that is frequently believed to be impossible in classical physics. Firstly, we have shown that orbits can be quantized in the phase space, producing one or more energy levels. We believe that the fact that these levels are produced in a finite number, as compared to having an infinite spectra of energy levels, is due to the fact that our delayed differential equations are not of the advanced type, as encountered in electrodynamics \cite{lop222}. Secondly, we have found sensitivity to initial conditions in the history space, what introduces unpredictability in a simple fashion, making the concept of randomness redundant, in principle \cite{spr07}. Are the apparent random fluctuations of fundamental physical systems just a byproduct of the complicated, even heterogeneous and high-dimensional \cite{sai21}, chaotic dynamics introduced by the dynamics of fields and the subsequent retardation effects in functional differential equations? \cite{lop222}. Finally, we have uncovered a robust intermittency in the absence of multistable external wells, simply caused by the inherent multiscale nature of our chaotic system. Of course, this is possible because retardation introduces more dimensions in the dynamical system, ultimately approaching its center or slow manifold. In this respect, a deep connection between Lorenz-like chaotic dynamical systems and walking droplets has been recently proved \cite{val22}.

Interestingly, other related phenomena commonly attributed to the microscopic realm, such as tunneling through external potential barriers (or in multistable external potentials) can be easily demonstrated with our retarded potential by introducing an external Duffing potential in replacement of the harmonic well used here \cite{coc18}. A similar situation occurs when studying the flow of electrons through potential barriers, where this paradoxical phenomenon becomes explained when interpreted in terms of the quantum potential, which appears in the Hamilton-Jacobi equation of the quantum system, and which is frequently disregarded when interpreting physical phenomena \cite{boh52}. For a connection between retarded potentials and the quantum potential we refer the reader to previous works \cite{lop20}. In other words, we are suggesting that the switch between different wells leading to an intermittent behavior can be interpreted in terms of the robust intermittency phenomenon. This dynamics is due to the nonlinear resonances that allow the particle to jump back and forth over the potential barrier \cite{coc18}.

Another important phenomena that might be studied with our oscillator is the existence of entangled states, which can be explained in terms of synchronization of oscillations \cite{lop20}. These states have already been predicted in previous works in classical electrodynamics to arise as a consequence of delay-coupling $\tau_i(x_i,x_j)$ and synchronization between systems of self-oscillating bodies. Synchronization phenomena has already found to actually produce entanglement in theoretical models of bouncing silicone oil droplets \cite{pap22}, although not with dynamical setups closing the locality loophole so far. Synchronization is more complicated for fluids, because the dissipation is higher at the scale of macroscopic fluid dynamics. Specially when compared to electrodynamic fields, where light travels mostly unimpeded when particles communicate through the electrovacuum. This can entail loopholes produced by the long-range correlations in the background fields \cite{mor06}.

Importantly, time-delays are frequently considered constant, so that their dynamical nature is disregarded. Fortunately, thanks to the development of numerical methods and computational techniques, an increasing number of works in the literature of dynamical systems is being dedicated to the dynamical evolution of time-delays \cite{mul18}. We have shown that the state-dependence of delays can produce very complicated behavior, entailing nonlinear oscillations through the ubiquitous Hopf bifurcation, and producing counterintuitive new complex dynamical chaotic behavior. The connection between state-dependent time-delayed differential equations and Li\'enard systems had been barely suggested \cite{jen13}. A much deeper exploration has been provided here. It was certainly lacking in the literature, and opens forefront possibilities to study new physical nonlinear phenomena.

In summary, we have provided new evidence in support of Raju-Atiyah's hypothesis, claiming that physical phenomena in the microscopic physical realm can be understood by using functional differential equations to study dynamical phenomena produced by time retardation in non-Markovian systems. Importantly, we highlight that the dissipation and the time-delay, which both constitute genuine radiative phenomena, introduce an arrow of time in physical systems \cite{mac03}. Thus perhaps the time-reversal symmetry of conservative field theories might be broken when oscillating and radiating solitons are formed in these fields \cite{fod06}. Partly, the abusive neglect of delayed feedback in physics stems from the tradition of Newtonian mechanics, where action at a distance is artificially introduced to simplify forces of interaction. Certainly, this approximation has rendered many accurate and valuable results, allowing a great progress in the knowledge of many macroscopic physical systems, which would have been impossible otherwise. Quite the opposite, the principle of causality in classical field theories produces memory effects that are always present whenever physical entities communicate through a background field with themselves, and among each other.

\section{Acknowledgment}

The author would like to thank Mattia Coccolo for valuable comments on the elaboration of the present manuscript, the discussion of some of its ideas and the computation of the basins of attraction.

\end{document}